\documentclass[prd,eqsecnum,preprint,showpacs,nofootinbib]{revtex4}
\usepackage{amsthm}
\usepackage{amssymb}
\usepackage{amsmath}
\usepackage{graphicx}

\newcommand{\be}{\begin{equation}}
\newcommand{\ee}{\end{equation}}
\newcommand{\bea}{\begin{eqnarray}}
\newcommand{\eea}{\end{eqnarray}}
\newcommand{\nn}{\nonumber}

\begin{document}

\title{Casimir Energies of Cylinders: Universal Function}
\author{E. K. Abalo}
\email{abalo@nhn.ou.edu}
\author{K. A. Milton}
\email{milton@nhn.ou.edu}
\affiliation{Homer L. Dodge Department of Physics and Astronomy, 
University of Oklahoma, Norman, OK 73019}
\author{L. Kaplan}
\email{lkaplan@tulane.edu}
\affiliation{Department of Physics, Tulane University, New Orleans,
LA 70118}
\date{\today}

\newcommand{\rperp}{\mathbf{r_{\bot}}}
\newcommand{\rperpp}{\mathbf{r_{\bot}^{\prime}}}

\begin{abstract}
New exact results are given for the interior Casimir energies of 
infinitely long waveguides of triangular  cross section 
(equilateral, hemiequilateral, and isosceles right triangles). Results for 
cylinders of rectangular cross section are rederived.   In particular, 
results are obtained for interior modes belonging to Dirichlet
and Neumann boundary conditions (TM and TE modes).  These results are 
expressed in rapidly convergent series using the Chowla-Selberg formula, 
and in fact may be given in closed form, except for general
rectangles.  The energies
are finite because only the first three heat-kernel coefficients
can be nonzero for the case of polygonal boundaries.
What appears to be a universal behavior of the Casimir energy as a 
function of the shape of the regular or quasi-regular
cross-sectional figure is presented. Furthermore, numerical calculations for
arbitrary right triangular cross sections suggest that the universal
behavior may be extended to waveguides of general polygonal cross sections. The new exact
and numerical results are compared with the proximity force approximation (PFA).

\end{abstract}

\maketitle

\section{Introduction}
Casimir's seminal 1948 paper \cite{casimir}, demonstrating that two 
perfectly conducting parallel plates placed in vacuum attract, created 
something of a stir in the physics community because of its unexpected
nature.  In principle, it was verified experimentally in 1958 by 
Sparnaay \cite{sparnaay}, although conclusive measurements had to wait
until the end of the 20th century \cite{lamoreaux, mohideen,Decca:2005yk}. 
Even more sensational was Boyer's 1969 \cite{boyer} calculation for the 
Casimir energy of a perfectly conducting spherical shell, which produced a 
repulsive force, rather than the attractive force that Casimir had postulated 
earlier \cite{casimir56} for a model of the electron. Since then, the question 
of how the sign and magnitude of the Casimir energy depend on the geometry 
of the system has been a constant leitmotif in Casimir energy studies.

In this paper, we attempt to elucidate this question by calculating the 
Casimir energies for the interiors of infinitely long cylinders of various 
cross sections. Such calculations have long been available for waveguides
of rectangular \cite{lukosz,ambjorn} and circular 
\cite{deraad,gosdzinsky}  
cross sections, but it apparently was not widely appreciated that energies for
certain triangular cross sections were equally well calculable.
We will work with triangular (equilateral, hemiequilateral, and isosceles 
right triangles), square, and rectangular cross sections. The triangular and 
rectangular cylinders are ideal since their eigenvalues are known explicitly. 
We obtain results in a very efficient form extremely amenable to numerical
evaluation by use of the Chowla-Selberg formula \cite{elizalde,cs,lerch}, 
which is far more
rapidly convergent than the direct evaluation of Epstein zeta functions.
Even better, we are able to obtain closed-form results for the interior
Casimir energies by use of a convenient representation found by
many authors
 \cite{lorenz,hardy,fletcher,zucker,gz,itzykson,itzykson2,kvitsinsky}.  
These powerful methods
supersede the previous method of directly summing the Epstein zeta
function. 
With the obtained data, we can then relate Casimir energies for these 
cross sections with their distinguishing geometrical properties, notably 
the perimeter $P$ and  the cross-sectional area $A$.  Our chief result
is the discovery that there appears to be universal functional
dependence for the Casimir energy per length multiplied by the area,
$\mathcal{E}A$,  due to the interior fluctuations inside an infinitely long 
waveguide, in terms of a dimensionless quantity expressing the waveguide's 
cross-sectional attributes, $A/P^2$.  This universality class applies to
figures based on regular polygons; for a rectangle, for example, of sides
$a$ and $b$ a different dependence obtains, which changes sign, as
expected, when the rectangle becomes sufficiently elongated.
 
It is interesting to note that the square and circular waveguides have been 
extensively studied \cite{lukosz,ambjorn,lukosz2,
lukosz3,zimerman,zimerman2,deraad, gosdzinsky, nesterenko}, 
in contrast to the triangular case which has only partially
 been examined. As far as 
the authors can tell, this paper is the first with explicit numerical results 
for the interior Casimir energy for infinitely long waveguides of triangular 
cross sections. Earlier work has, however, been done on plane
equilateral triangular geometries \cite{inui}.  For other work on equilateral
triangular domains see Refs.~\cite{itzykson,itzykson2,kvitsinsky,
ahmedov,hazlett}.
There has been recent work in which
explicit eigenvalues for the equilateral triangle have been used in
computing numerical values for a piston of such a cross section \cite{alvarez}.

Probably the reason why these Casimir energies 
were never evaluated, although the 
explicit eigenvalues appear in textbooks \cite{embook,radbook}, is that
most authors considered interior calculations suspect.  It is, of course,
not possible by these methods to extract exterior eigenvalues or the
corresponding Casimir energies.  For smooth surfaces, such as a circular
cylinder, only the sum of the interior and exterior Casimir energies of
an infinitesimally thin boundary shell can be calculated unambiguously.
In the language of heat kernels, the $a_2$ heat kernel coefficient is
proportional to the cube of the
curvature, and cancels only when both interior and
exterior modes are included.  When $a_2\ne0$ and only the
interior is considered, there is a logarithmic
divergence that cannot be removed, leaving an ambiguous Casimir energy.
But, for polygonal boundaries, there is no such coefficient, and the
interior Casimir energies seem well defined.  There are divergences 
associated with the corners, but
they  do not contribute to the global Casimir energy
of the waveguide, because they may be unambiguously subtracted off.  
For a review of the state of the art for self-energies
of smoothly bounded regions, see Ref.~\cite{Milton:2010qr}

The outline of this paper is as follows.  In Sec.~\ref{sec2} we formulate
Casimir energies in terms of mode summations, which we regulate either
by dimensional continuation or by a temporal point-splitting cutoff.
The former approach leads naturally to the use of the Chowla-Selberg formula,
while the latter produces an expression that is summed through use of the
Poisson summation formula; either approach can lead to a result that can,
in special cases, be expressed in closed form.
These machineries are applied to cylinders with equilateral triangular
cross sections in Sec.~\ref{sec3}, for both Dirichlet and Neumann scalar
modes, and hence for interior electromagnetic modes for a perfectly
conducting boundary.  The same is done in Sec.~\ref{sec4} for a
bisected equilateral triangular cross section, that is, a hemiequilateral
or 30$^\circ$-60$^\circ$-90$^\circ$ triangle.  A square cylinder is
reconsidered in Sec.~\ref{sec5}, as is a rectangular cylinder in 
Sec.~\ref{sec6}. The bisection of a square waveguide, one formed by a
right isosceles triangle, is treated in Sec.~\ref{sec7}. 
General right triangular cross sections are considered
numerically in Sec.~\ref{sec8}, showing that a finite Casimir energy is
obtained even when the eigenvalues are not explicitly available.  All
the energies for triangular  Dirichlet boundaries found in this paper
lie on a universal curve, which smoothly joins the proximity force
approximation. Why the interior modes of a continuous curved
cross section, such as that of a circle, cannot yield a finite Casimir
energy is discussed in Sec.~\ref{sec9}.  The corner divergences are
reflected only in the $a_1$ heat-kernel coefficient, unlike the curvature
divergences which show up also in the $a_2$ coefficient.  Again, unlike
for continuously curved figures in a plane,  Casimir energies for squares and
equilateral triangles, and figures obtained from these by bisection,
are likewise unambiguously calculable, and given in Sec.~\ref{sec10},
along with numerical results for right triangles.
Results for the cylindrical geometries considered are summarized 
in the Conclusions in Sec.~\ref{sec11}.  The Appendix sketches the derivation of the
Chowla-Selberg summation formula from the Abel-Plana formula.

\section{Casimir Energy of a Cylinder}\label{sec2}
	\subsection{Mode summation}
The Casimir energy per length, $\mathcal{E}$, 
for infinite cylindrical geometries can generally be expressed as a 
sum over mode frequencies in the following form,
\begin{equation}\label{CasEgeneral}
\mathcal{E}=\frac{1}{2} \int_{-\infty}^{\infty} \frac{dk}{2\pi}\,
\underset{m,n} {\sum} \sqrt{k^2+ \gamma_{m n}^{2}}\quad,
\end{equation}
where $\gamma_{m n}^2$ are the eigenvalues of the two-dimensional Laplacian, 
$\left(\nabla^2_{{\bot}}+\gamma^2_{m n}\right)\Phi_{m n}(\rperp)=0$, subject
to $\Phi_{m n}(\rperp)$ satisfying appropriate boundary conditions on the 
cylinder's surface. Here, $\rperp$ denotes the two coordinates transverse to 
the longitudinal cylinder axis. For cases where the eigenvalues have explicit 
formulas, $\gamma_{m n}^2$ is a quadratic function of the mode numbers $m$ 
and $n$. 

The form of this quadratic function depends on the geometry, and
the range of the mode numbers depends on the boundary conditions being applied.
In the subsequent sections, we will look at the cases of Dirichlet and Neumann 
boundary conditions. The electromagnetic case can be obtained from these two
classes of modes. Because of the cylindrical geometry, the modes of the 
electromagnetic field in a perfectly conducting waveguide can be expressed in 
terms of two scalar fields 
satisfying separately Dirichlet (E or TM modes) and Neumann boundary 
conditions (H or TE modes) \cite{embook,radbook}. It follows then that the 
Casimir energy for an electromagnetic field is simply the sum of the Casimir 
energies for the E and H modes. (This breaks down if the boundaries are
imperfect, for example, for a general dielectric cylinder \cite{stratton}.)

The expression above for the energy is formally divergent and as 
usual requires a regularization technique in order to obtain physically 
interpretable results. Out of the many ways to regularize such an expression, 
we choose the methods of dimensional regularization and point-splitting 
regularization.
	
\subsection{Regularization methods}
\subsubsection{Dimensional regularization}
By extending the one dimension for the longitudinal wavevector
$k$ to $d$ dimensions, $dk \rightarrow d^{d}k$, we obtain the following 
expression for the energy per length,
\begin{equation}\label{CasEDimReg}
\mathcal{E}=-\lim_{d\to1}\frac{\Gamma(-(1+d)/2)}{2^{d+2}\pi^{(d+1)/2}}
 \sum_{m,n} \left( \gamma_{m n}^2\right)^{(1+d)/2}.
\end{equation}
In the limit $d \rightarrow 1$, we obtain a $\Gamma(-1)$ divergence. 
However, one can surmount this problem by analytic continuation. For the
cases considered in this paper, the eigenvalue expression $\gamma_{m n}^2$ 
is a simple quadratic form in $m$ and $n$. This allows us to employ the 
Chowla-Selberg formula,  an exact formula for a class of 
Epstein zeta functions \cite{elizalde,cs}, proved in the Appendix,  
and thereby utilize the reflection property of the zeta function, 
\be\Gamma(s)\zeta(2s)=\Gamma((1-2s)/2)\zeta(1-2s)\pi^{2s-1/2},\label{refl}
\ee which allows
us to continue to $d\to1$.
The technique of dimensional regularization is our method of choice throughout 
this paper since the Chowla-Selberg formula converges exponentially faster 
than the Epstein zeta functions obtained by the point-splitting 
regularization we present below. 

In fact, it has been known for many years that we can 
sum the zeta function into a form 
\cite{lorenz,hardy,fletcher,zucker,gz,itzykson,itzykson2} 
that can yield a closed-form expression for the Casimir
energy in the cases of a square, isosceles right triangle,
equilateral right triangle, and a 30$^\circ$-60$^\circ$-90$^\circ$
triangle.
		
\subsubsection{Point-splitting regularization (Cutoff)}
A more physical approach is to start from the general form for
the quantum vacuum energy in terms of the Green's function for a given
frequency $\omega$,
\bea
E=\frac1{2i}\int(d\mathbf{r})\int\frac{d\omega}{2\pi}2\omega^2\mathcal{G}
(\mathbf{r,r}),
\eea
which for the situation of a cylindrical waveguide, gives the
energy per length
\be
\mathcal{E}
=\frac12\int_{-\infty}^\infty \frac{d\zeta}{2\pi}2(-\zeta^2)\int
\frac{dk}{2\pi}\sum_{m,n}\frac1{\zeta^2+k^2+\gamma_{mn}^2}e^{i\zeta\tau},
\ee
where in the second line we have made a Euclidean rotation,
$\omega\to i\zeta$, as well as introduced a time-splitting regulator,
$t-t'\to i\tau$, where we are to take $\tau$ to zero at the end of the
calculation.  Assuming $\tau>0$, we do the $\zeta$ integration by closing
it in the upper half plane, with the result
\bea
\mathcal{E}&=&\frac12\lim_{\tau\to0}\sum_{m,n}
\int_{-\infty}^\infty \frac{dk}{2\pi}
\sqrt{k^2+\gamma_{mn}^2}e^{-\tau\sqrt{k^2+\gamma_{mn}^2}}\nn\\
&=&\frac{1}{2}\, \lim_{\tau\to0} \left(- \frac{d}{d\tau} \right) 
\int_{-\infty}^{\infty} \frac{dk}{2\pi}\,
\sum_{m,n} e^{-\tau \sqrt{k^2+ \gamma_{m n}^{2}}}.\label{e-ptsplt}
\eea

Just as the dimensional regularization method was paired with the analytic 
continuation of the exponent, followed by use of the Chowla-Selberg
formula, the cutoff regularization method is combined 
with Poisson's sum formula. After rearranging the summations to span 
$-\infty$ to $+\infty$, the summation is converted by Fourier transformation
from one involving powers of $e^{-\tau}$ to powers of $e^{-1/\tau}$.
The final expression after simplifications is similar to the 
Chowla-Selberg result except for a slowly converging double sum. As noted
above, in fact, that double sum can in special cases
 be summed into an explicit closed
form; both regularizations give the same answer 
and thus serve as somewhat 
independent consistency checks. In addition, the cutoff regularization, 
unlike the dimensional-zeta-function 
regularization, exhibits the explicit divergences. 
We are then able to identify a ``volume" and a ``surface" divergence, 
as well as a divergent term independent of the
scale of the cross section. All of these
divergent terms can be regarded as contact
terms, and, therefore, may be discarded.  Corner divergences occur only
in the constant term.

\section{Equilateral Cylinder}\label{sec3}
\begin{figure}
\begin{center}
\includegraphics[scale=0.4]{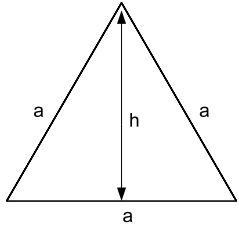}
\caption{\label{fig:eqt} Equilateral triangle of side $a$ and height $h=
\sqrt{3} a/2$.}
\end{center}
\end{figure}

We consider an infinitely long cylinder of equilateral triangular cross 
section with side length $a$ and height $h$, as shown in Fig.~\ref{fig:eqt}. 
The eigenvalues of the Laplacian on an equilateral triangular surface 
with Dirichlet or Neumann boundary conditions (solved by G. Lam\'e \cite{lame}) are of the form  
$\gamma_{m n p}^{2}=\frac{2 \pi^2}{3 h^2}(m^2+n^2+p^2) $, where 
$m+n+p=0 $ \cite{embook,radbook}. Using this constraint, the previous 
expression can be simplified resulting in 
\be
\gamma_{m n}^{2}=\frac{4 \pi^2}{3 h^2}(m^2+ m \, n+n^2).\label{gamma}
\ee
Here, $m$ and $n$ are positive or negative integers; their allowed values
depend on the boundary conditions, as indicated
below. 
The eigenfunctions (which are given
in Refs.~\cite{embook, radbook}) have degeneracy under cyclic or anticylic
permutations of the three indices $m$, $n$, $p$, so a factor of $1/6$ must
be inserted before the mode sum.
One would, indeed, expect only two mode numbers as it is a two-dimensional 
surface; however, the earlier form is more useful for counting degeneracies.

\subsection{Dirichlet boundary}

Dirichlet boundary conditions require the fields to vanish on the 
boundary $C$, $\Phi({\rperp})|_C=0 $. The mode numbers are now restricted to 
$m \neq 0 $, $n \neq 0 $, and $p \neq 0 $. If any index vanishes,
the corresponding eigenfunction would likewise vanish. One can then use 
Eq.~$\eqref{CasEgeneral}$ with Eq.~(\ref{gamma}), and perform the appropriate
sum over $m$ and $n$ to obtain the Casimir energy. The dimensionally regulated
expression  $ \eqref{CasEDimReg}$ may be evaluated by
using the Chowla-Selberg formula, for $\mbox{Re\,} s>1$:
\begin{eqnarray}
&&\sum_{m,n=-\infty}^\infty\!\!\!\!{}^{\prime\prime}(am^2+b m n+ c n^2)^{-s}  = 
2a^{-s}\zeta(2s)+\frac{2^{2s}\sqrt{\pi}a^{s-1}}{\Gamma(s)\Delta^{s-1/2}}
\zeta(2s-1)\Gamma(s-1/2)\nn\\
& &\quad\mbox{}+\frac{2^{s+5/2}\pi^{s}}{\Gamma(s)\Delta^{s/2-1/4}\sqrt{a}}
\sum_{n=1}^\infty n^{s-1/2}\sigma_{1-2s}(n)\cos(n\pi b/a) 
K_{1/2-s}(n \pi \sqrt{\Delta}/a).\label{cs}
\end{eqnarray}
However, it is valid only when the discriminant $\Delta=4ac-b^2 > 0 $. 
The double
prime indicates $m=n=0$ is excluded from the summation range. The divisor 
function, $\sigma_{k}(n)$, is the sum of the $k$-th powers of the divisors of 
$n$,
\begin{equation}
\sigma_{k}(n)\equiv \underset{d|n}{\sum}d^{k}.
\end{equation}

In the particular case of the Dirichlet equilateral triangle, $\Delta=3 $. 
The special mode sum to be evaluated involves the explicit exclusion of
the case when one of the mode labels vanishes:
\bea
\mathcal{E}^{(D)}_{\rm Eq}
&=&-\frac{\Gamma(-(1+d)/2)}{2^{d+2}\pi^{(d+1)/2}}\left[\frac43\left(
\frac\pi h\right)^2\right]^{(d+1)/2}\nn\\
&&\quad\times\frac16\left\{
\sum_{m,n}{}''(m^2+n^2+mn)^{(1+d)/2}-6\zeta(-1-d)\right\}.
\eea
After a few manipulations such as use of the reflection property 
(\ref{refl}) we obtain
\bea
\mathcal{E}_{\rm{Eq}}^{(D)} & = & \frac{1}{144 \pi^{2} h^2}
\left( 
8 \pi \zeta(3)- 3^{3/2}\zeta(4)-4\pi^2(12)^{3/4}\sum_{n=1}^\infty 
n^{-3/2}(-1)^{n}\sigma_{3}(n)K_{3/2}(n \pi \sqrt{3}\,)
\right).\nn\\
\eea
The above expression converges very fast and reaches an accuracy of twenty 
decimal places when just the first eight terms in the $n$ sum are included,	
\begin{equation}
\mathcal{E}_{\rm{Eq}}^{(D)}= \frac{0.017789138469130117062}{h^2}.
\label{eq-dirichlet}
\end{equation}

\subsection{Neumann boundary}
For Neumann boundary conditions, the normal derivative of the eigenfunctions 
must vanish on the boundary, $ \partial_n \Phi({\rperp})|_C=0$. The 
restriction on the mode numbers is less severe than the Dirichlet case:
here one of the mode numbers can be zero, but not two simultaneously. 
This gives a different summation range from that for the Dirichlet case, 
and after similar regularization manipulations, we obtain the result
\begin{subequations}
\bea
\mathcal{E}_{\rm{Eq}}^{(N)}&=&
\mathcal{E}_{\rm{Eq}}^{(D)}-\frac{\zeta(3)}{6 \pi h^2}\label{e-eq-n1}\\
&=&-\frac{0.045982}{ h^2}.
\eea
\end{subequations}

\subsection{EM perfectly conducting boundary}
As discussed in the Introduction, the electromagnetic Casimir energy in the 
interior of perfectly conducting waveguide having an equilateral triangular 
cross section is simply the sum of the two previous energies,
\begin{subequations}
\bea
\mathcal{E}_{{\rm Eq}}^{({\rm EM})}&=&\mathcal{E}_{{\rm Eq}}^{(D)}+
\mathcal{E}_{{\rm Eq}}^{(N)}\label{em-eq1}
\\
&=&-\frac{0.028193}{ h^2}.\label{em-eq}
\eea
Note that while the Dirichlet energy is positive, the electromagnetic energy
is negative, because the H mode overwhelms the E mode.
\end{subequations}

\subsection{Point-splitting regularization}
Poisson's summation formula gives an expression equivalent to the original sum in terms 
of the summand's Fourier transform:
\begin{equation}
\sum_{p=-\infty}^\infty f(p)=\underset{q=-\infty}{\overset{\infty}{\sum}}
\left(\int_{-\infty}^{\infty} d p \,e^{2\pi i p \, q}  f(p)
\right).\label{psf}
\end{equation}	
This is useful for converting formulas most readily applicable for a large
parameter to one most convenient when the parameter is small---that is,
it is a duality transformation.
	
We will demonstrate this regularization method with the Dirichlet 
equilateral triangle. The mode sum in the energy (\ref{e-ptsplt})
can be given explicit form as
\begin{equation}
\sum_{m,n}e^{-\tau \sqrt{k^2+\gamma_{mn}^2}}=\frac{1}{6}
\left(\sum_{m,n=-\infty}^{\infty} e^{-\tau \sqrt{k^2+\gamma_{mn}^2}}
-3 \, \sum_{m}e^{-\tau \sqrt{k^2+\gamma_{m0}^2}}
+2\,e^{-\tau \sqrt{k^2}}   \right),
\end{equation}
where $\gamma_{mn}^2$ is given by Eq.~(\ref{gamma}), and the sums on the
right hand side extend over all integer values, positive, negative, and zero. 
We first perform a transformation on $\gamma_{mn}$ with 
$m=r+s$ and $n=r-s$.
It follows that
\begin{eqnarray}
\sum_{m,n}e^{-\tau \sqrt{k^2+\gamma_{mn}^2}} & \rightarrow &
\frac{1}{6}\bigg\{ \sum_{r,s} \left[ e^{-\tau \sqrt{k^2+\gamma_{rs}^2}}+
e^{-\tau \sqrt{k^2+\gamma_{r+1/2,s+1/2}^2}} \right] \nonumber\\
& &\quad\mbox{}-3 \, \sum_r(e^{-\tau \sqrt{k^2+\gamma_{0r}^2}})
+2 e^{-\tau \sqrt{k^2}}   \bigg\},  
\end{eqnarray}
where now $\gamma_{rs}^2=\frac{4 \pi^2}{3 h^2}(3r^2+s^2) $,
and the $r$ and $s$ sums again extend over all integers.
In a manner similar to Ref.~\cite{lukosz}, we adopt spherical coordinates and 
evaluate the generic Fourier transforms:
\bea
&&\int_{-\infty}^{\infty} dk\, \sum_{r,s}{e^{-\tau \sqrt{k^2+\beta \lbrace 
a(r+b)^2+c(s+d)^2 \rbrace}}}\nn\\
&=&\sum_{p,q}e^{-2\pi i b p}e^{-2\pi i d q}
\int dk\,dr\,ds\,e^{2\pi i rp}e^{2\pi i s q}
e^{-\tau\sqrt{k^2+\beta(ar^2+cs^2)}}\nn\\
&=&\sum_{p,q}e^{-2\pi i b p}e^{-2\pi i d q}
\frac1{\beta\sqrt{ac}}\int_0^\infty R^2\,dR\int_0^\pi \sin\theta\,d\theta
\int_0^{2\pi}d\phi e^{i\mathbf{R\cdot p}}e^{-\tau R},
\eea
where we have defined the vectors $\mathbf{R}=
(\sqrt{\beta a}r,\sqrt{\beta c}s,k)$ and 
$\mathbf{p}=\frac{2\pi}{\sqrt{\beta}}(p/\sqrt{a},q/\sqrt{c},0)$.
Carrying out the radial and angular integrals, we find straightforwardly
\begin{subequations}
 \bea
\int_{-\infty}^{\infty} dk\, \sum_{r,s}e^{-\tau \sqrt{k^2+\beta \lbrace 
a(r+b)^2+c(s+d)^2 \rbrace}}
&=&
\sum_{p,q}\frac{e^{-2 \pi i(bp+dq)}\beta}{\sqrt{a c}}\frac{8 \pi \tau}
{\left[ \tau^2 \beta +4 \pi^2 (p^2/a+q^2/c)\right]^2},\nn\\ \\
\int_{-\infty}^{\infty} dk\, \sum_r{e^{-\tau \sqrt{k^2+\beta a r^2}}}
&=&
\sum_p\frac{\beta}{\sqrt{a}}\frac{2 \pi \tau}{\left( \tau^2 \beta 
+4 \pi^2 p^2/a\right)^{3/2}},\\
\int_{-\infty}^{\infty} dk\,e^{-\tau \sqrt{k^2}}
&=&\frac{2}{\tau},
\eea
\end{subequations}
in terms of the abbreviation $\beta=\frac43\left(\frac\pi{h}\right)^2$.
With the appropriate coefficients for the Dirichlet equilateral triangle case, 
we isolate the divergences in the energy:
\begin{equation}
\mathcal{\widehat{E}}_{\rm Eq}^{(D)}
=\lim_{\tau \to 0}\left(\frac{\sqrt{3} \, h^2}{2 \pi^2 \tau^4}-
\frac{\sqrt{3} \, h}{4 \pi \tau^3}+\frac{1}{6\pi \tau^2}\right) 
=\lim_{\tau\to0}\left(\frac{3A}{2\pi^2\tau^4}-\frac{P}{8\pi\tau^3}+\frac{C}{48\pi\tau^2}
\right).\label{div-e}
\end{equation}
We note that the ``volume" and ``surface" divergent terms,
which are respectively proportional to 
the area of the triangle $A=h^2/\sqrt{3}$  and the 
perimeter $P=2\sqrt{3}h$, are as expected,
and are presumably not of physical relevance.  
The last term, a constant in $h$, certainly does
not contribute to the self-stress on the cylinder. Only this term reflects
the corner divergences, and for the equilateral triangle we have $C=8$.  For a general polygon, with interior angles
$\alpha_i$, the corner coefficient is
\be
C=\sum_i \left(\frac\pi{\alpha_i}-\frac{\alpha_i}\pi\right).
\label{cornerterm}
\ee  The coefficients in Eq.~(\ref{div-e}) are proportional to the heat kernel coefficients
discussed in Sec.~\ref{circular} below.

After these terms are isolated, the finite part of the energy is given by
the following expression:
\begin{equation}
\mathcal{E}_{\rm Eq}^{(D)}=\frac{1}{144 \pi^2 h^2}
\left(12\pi\zeta(3)-\frac{10\sqrt{3}}{3}\zeta(4)-\frac{16\sqrt{3}}{3}
\sum_{p,q=1}^\infty\frac{1+(-1)^{p+q}}{(p^2+q^2/3)^2}
\right).\label{alt-eq}
\end{equation}
	
The double sum in the above expression converges slowly. Summing the first 500 terms for $m$ and $n$, we reach a seven-decimal accuracy,
\begin{equation}
\mathcal{E}_{{\rm Eq}}^{(D)}= \frac{0.0177891}{h^2},
\end{equation}
which agrees with Eq.~(\ref{eq-dirichlet}).  
Because of the numerical agreement,
one is convinced that both regularization methods yield 
exactly the same answer and we have thus shown that 
our calculations are correct.

A virtue of the alternative form (\ref{alt-eq}) is that it is now
evident from Eq.~(\ref{e-eq-n1}) 
that in the electromagnetic energy (\ref{em-eq1}) the $\zeta(3)$ term 
completely cancels.  As a result, the electromagnetic energy is manifestly
negative.  This is a feature that will persist in all the examples treated
in this paper.

\subsection{Closed-form result}
Remarkably, for the integrable polygonal figures we are considering,
the Casimir energy can be given in closed form.  Following 
Refs.~\cite{fletcher,zucker,gz,itzykson,itzykson2}, 
we write for the equilateral triangular cross section, from 
Eq.~(\ref{CasEDimReg}),
\be\mathcal{E}_{\rm Eq}^{(D)}
=-\lim_{s\to-1}\frac12(4\pi)^s\Gamma(s)\left(\frac{4\pi^2}{3h^2}
\right)^{-s}\left[\zeta(s)L_3(s)-\zeta(2s)\right],
\ee
in terms of the single series
\be
L_3(s)=\sum_{n=0}^\infty\left[\frac1{(3n+1)^s}-\frac1{(3n+2)^s}\right].
\ee
Now, this function satisfies the reflection property
\be
L_3(s)\Gamma(s)=\frac{\sqrt{3}(2\pi/3)^s}{2\sin s\pi/2}L_3(1-s).
\ee
Then, using Eq.~(\ref{refl}), we can take the limit $s\to-1$:
\be
\mathcal{E}=-\frac1{96h^2}\left[\sqrt{3}L_3(2)-\frac8\pi \zeta(3)\right].
\ee
In fact, the remaining sum has a closed form:
\be
L_3(2)=\frac19\left[\psi'(1/3)-\psi'(2/3)\right],
\ee
in terms of the polygamma function.  Thus
\be
\mathcal E_{\rm Eq}^{(D)}=-\frac1{96 h^2}\left[\frac{\sqrt{3}}9\left[
\psi'(1/3)-\psi'(2/3)\right]-\frac8\pi \zeta(3)\right]=\frac{0.0177891}{h^2}.
\ee
It is {\it a priori\/} remarkable that such an explicit form can be achieved
for a strong-coupling problem.

In particular, from Eqs.~(\ref{e-eq-n1}) and (\ref{em-eq1}), 
we see that the interior
Casimir energy for a perfectly conducting cylinder with equilateral
triangular cross section has the simple form
\be
\mathcal{E}_{\rm Eq}^{({\rm EM})}=-\frac{\sqrt{3}}{432h^2}[\psi'(1/3)
-\psi'(2/3)],
\ee that is, as we noted above, the $\zeta(3)$ cancels.

\section{Hemiequilateral Cylinder}\label{sec4}
\begin{figure}
\begin{center}
\includegraphics[scale=0.4]{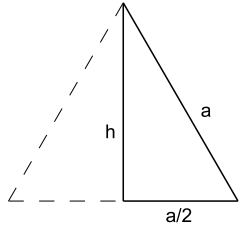}
\caption{\label{fig-369}$30^\circ$-$60^\circ$-$90^\circ$ 
triangle obtained by bisecting an equilateral triangle.}
\end{center}
\end{figure}

The hemiequilateral triangle or $30^\circ$-$60^\circ$-$90^\circ$ triangle 
is appropriately named 
since it is obtained by bisecting the equilateral triangle, 
as shown in Fig.~\ref{fig-369}. 
Hence, taking an equilateral triangle of side length $a$ and 
height $h$ we obtain a 
hemiequilateral triangle of height $h$ and side lengths $a$ and $a/2$. 
This relation between 
the two triangles proves very useful in realizing that the hemiequilateral 
modes are a 
subset of the equilateral modes \cite{radbook}.  
(The proof that there are no
additional modes is given in that reference.)  Therefore, one 
simply has to select 
the equilateral modes that satisfy chosen boundary conditions 
on the bisector. 
It is implied that the eigenvalues are of the same form as the equilateral 
triangle's, Eq.~(\ref{gamma}).

\subsection{Dirichlet boundary}
To satisfy Dirichlet boundary conditions for the hemiequilateral triangle, 
the eigenmodes 
must vanish on all three sides of the hemiequilateral triangle. The Dirichlet 
modes 
for the equilateral triangle already satisfy that condition on two sides, 
so we are 
to select the ones that vanish on the bisector. After a few manipulations, 
the Casimir 
energy for the Dirichlet hemiequilateral triangle is related to the 
equilateral Dirichlet result with the formula, 
\begin{equation}
\mathcal{E}_{369}^{(D)}=\frac{\mathcal{E}_{\rm Eq}^{(D)}}{2}
+\frac{\zeta(3)}{8 \pi h^2}
=\frac{0.0567229}{h^2},\label{d-hemi}
\end{equation}
a positive energy again.
\subsection{Neumann boundary}
Similarly, the Neumann modes for the hemiequilateral triangle 
have to be chosen from the Neumann 
equilateral modes such that their normal derivative on the bisector vanishes. 
With this new restriction on the mode numbers, the relation between Neumann 
hemiequilateral and equilateral Casimir energies is established,
\begin{equation}
\mathcal{E}_{369}^{(N)}=\frac{\mathcal{E}_{\rm{Eq}}^{(N)}}{2}
-\frac{\zeta(3)}{8 \pi h^2} 
=-\frac{0.0708193}{h^2},\label{n-hemi}
\end{equation}
which is again negative.	
\subsection{EM perfectly conducting boundary}
Straightforwardly, we obtain the electromagnetic Casimir energy 
in the interior of 
an infinitely long cylinder of hemiequilateral triangular cross section by 
adding the E and H modes,
\begin{equation}
\mathcal{E}_{369}^{({\rm EM})}=\mathcal{E}_{369}^{(D)}
+\mathcal{E}_{369}^{(N)}=\frac12\mathcal{E}^{({\rm EM})}_{\rm Eq}
=-\frac{0.0140964}{h^2},
\end{equation}
which is, remarkably, exactly one-half that of the energy of equilateral
triangle, as might have been anticipated naively.

\section{Square Cylinder}\label{sec5}
\begin{figure}
\begin{center}
\includegraphics[scale=0.4]{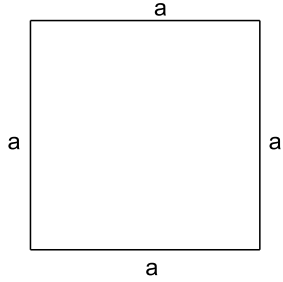}
\caption{Square of side $a$.\label{fig:sq}}
\end{center}
\end{figure}

For completeness, 
we consider a square of side length $a$, as illustrated
in Fig.~\ref{fig:sq}. Of course, this is the geometry originally
considered by Lukosz \cite{lukosz}, and by Ambj\o rn and Wolfram
\cite{ambjorn}. The form for the eigenvalues is 
$\gamma_{m n}^2=\frac{\pi^2}{a^2}(m^2+n^2)$ \cite{embook,radbook}. 
It is of a simpler quadratic form than the previous two geometries, 
which makes the counting of degeneracies somewhat simpler.
	
\subsection{Dirichlet boundary}
The Dirichlet modes for the square must vanish on all four sides, and so 
are products of two sine functions, a classic result in most electricity 
and magnetism books. The eigenvalues must therefore both be positive, 
$m > 0$ and $n > 0$. With the appropriate summation range and dimensional 
regularization, we obtain a result consistent with that
found in Refs.~\cite{lukosz, ambjorn}, 
where we also display the result found by the alternative cut-off method:
\begin{subequations}
\bea
\mathcal{E}^{(D)}_{\rm Sq}&=&-\frac1{32\pi^2a^2}\bigg[2\zeta(4)-\pi\zeta(3)
+8\pi^2\sum_{l=1}^\infty l^{-3/2}\sigma_3(l)K_{3/2}(\pi l)
\bigg]\label{cs-sq}\\
&=&-\frac1{32\pi^2a^2}\bigg[4\zeta(4)-2\pi\zeta(3)
+4\sum_{k,l=1}^\infty\frac{1}{(k^2+l^2)^2}\bigg]\\
 &=& 0.00483155/a^2\eea
\end{subequations}
The Chowla-Selberg formula (\ref{cs-sq})  is extraordinarily convergent.

Again this result can be given in closed form.  Actually, we can do this
directly from the double sum \cite{lorenz,hardy,zucker},
\be
\sum_{k,l=1}^\infty\frac1{(k^2+l^2)^2}=\zeta(2)L_4(2)-\zeta(4),
\ee
where 
\be
L_4(2)=\sum_{m=0}^\infty \frac{(-1)^m}{(2m+1)^2}=G,
\ee
where $G=0.915966\dots$ is Catalan's constant. Then
\be
\mathcal{E}_{\rm Sq}^{(D)}=\frac1{16\pi a^2}\left[\zeta(3)-\frac\pi3 G\right]=
\frac{0.00483155}{a^2}.
\ee

\subsection{Neumann boundary}
The Neumann modes for the square must have their normal derivative vanish on 
the boundary. Such modes are the products of two cosine functions, 
which implies that $m$ and $n$ cannot both be zero. 
The Neumann energy in relation to the Dirichlet energy is
\begin{equation}
\mathcal{E}_{\rm Sq}^{(N)}=\mathcal{E}_{\rm Sq}^{(D)}
-\frac{\zeta(3)}{8 \pi a^2}  =-\frac{0.0429968}{a^2},
\end{equation}
which again reproduces the known result.
\subsection{EM perfectly conducting boundaries}
Summing Dirichlet and Neumann energies, we obtain the following 
electromagnetic Casimir energy for the interior of a 
perfectly conducting square waveguide, expressed only in
terms of one transcendental number $G$:
\begin{equation}
\mathcal{E}_{\rm Sq}^{({\rm EM})}=\mathcal{E}_{\rm Sq}^{(D)}
+\mathcal{E}_{\rm Sq}^{(N)}=-\frac{G}{24 a^2}=-\frac{0.0381653}{a^2}.
\end{equation}

\section{Rectangular cross section}\label{sec6}
	
The more general expression for a Dirichlet rectangle is easily obtained
by the above methods.  Again we display the result in the two forms,
\begin{subequations}
\bea
\mathcal{E}_{\rm Rect}^{(D)}&=&
-\frac1{32\pi^2 ab}\left[-\pi\frac{b}a\zeta(3)+2\zeta(4)+
8\pi^2\left(\frac{b}a\right)^{1/2}\sum_{l=1}^\infty l^{-3/2}\sigma_3(l)
K_{3/2}(2\pi l b/a)\right]\\
&=&\frac{1}{32 \pi^2 a^2}\left[ \left(1+\left(\frac{a}b\right)^{2}\right)
\pi\zeta(3)
-2\left(\left(\frac{a}b\right)^{3}+\frac{b}a\right)\right.
\zeta(4)\nn\\
&&\qquad\mbox{}-4
\left.\left(\frac{a}b\right)^{3}\sum_{m,n=1}^\infty\left(m^2
+\left(n \frac{a}b\right)^{2}\right)^{-2}\right].\label{rec2}
\eea
\end{subequations}
For $a=b$, a square, the Dirichlet energy is positive, but when one side 
is much
larger than the other, the sign of the energy must change, for that situation
corresponds to the classic case of Casimir attraction.  Indeed, the 
Dirichlet Casimir energy
vanishes for $b/a=1.74437$, and is negative for larger values of $b/a$.  See 
Fig.~\ref{fig:rec}
below.  Note that for a general rectangle, $a\ne b$, a closed-form expression
for the energy is apparently not achievable \cite{gz}.

The corresponding Neumann result is simply
\begin{eqnarray}
\mathcal{E}_{\rm Rect}^{(N)} & = & \mathcal{E}_{\rm Rect}^{(D)} 
-\left(1+\left(\frac{a}b\right)^{2}\right)\frac{\zeta(3)}{16 \pi a^2},
\end{eqnarray}
which again shows from Eq.~(\ref{rec2}) that the electromagnetic Casimir
energy is always negative.

\section{Isosceles right cylinder}\label{sec7}
\begin{figure}
\begin{center}
\includegraphics[scale=0.4]{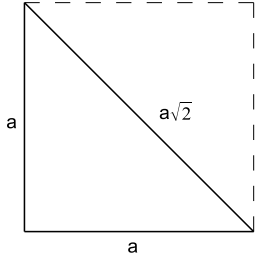}
\caption{\label{fig:iso} Isosceles right triangular waveguide, of side $a$.}
\end{center}
\end{figure}

The relation of the isosceles right triangle to the square is analogous 
to that of the 
hemiequilateral triangle to the equilateral triangle. 
Bisecting a square produces an 
isosceles right triangle, as seen in Fig.~\ref{fig:iso}. 
The eigenmodes of the isosceles right triangle are a subset 
of the square eigenmodes satisfying the extra condition 
that they obey given boundary 
conditions on the corresponding square's diagonal.  
The eigenfrequencies are therefore 
of the same form as the square's,
$\gamma_{m n}^2=\frac{\pi^2}{a^2}(m^2+n^2)$, with different ranges for 
$m$ and $n$ \cite{embook,radbook}.

\subsection{Dirichlet boundary}
Dirichlet boundary conditions on the isosceles right triangle 
translate into the constraint 
$m > n$, and $n \geq 1$. Just as in previous cases, the result 
follows seamlessly. 
The Casimir energy can be related to that of the Dirichlet square,
\begin{equation}
\mathcal{E}_{\rm Iso}^{(D)}=\frac{\mathcal{E}_{\rm Sq}^{(D)}}{2}
+\frac{\zeta(3)}{16 \pi a^2}=\frac{0.0263299}{a^2},
\end{equation}
a positive number again.
\subsection{Neumann boundary}
By imposing Neumann boundary conditions, the mode numbers must now satisfy  
$m \geq  n$, and $n \geq 0$. The relation to the Neumann case for the 
square is then,
\begin{equation}
\mathcal{E}_{\rm Iso}^{(N)}=\frac{\mathcal{E}_{\rm Sq}^{(N)}}{2}
-\frac{\zeta(3)}{16 \pi a^2}=-\frac{0.0454125}{a^2},\end{equation}
again a negative number.	
	
\subsection{EM perfectly conducting boundary}
We combine both results to obtain the electromagnetic Casimir energy for 
a perfectly conducting waveguide of right isosceles triangular cross section,
\begin{equation}
\mathcal{E}_{\rm Iso}^{({\rm EM})}=\mathcal{E}_{\rm Iso}^{(D)}+
\mathcal{E}_{\rm Iso}^{(N)}=\frac12 \mathcal{E}_{\rm Sq}^{({\rm EM})}
=-\frac{G}{48a^2}
=-\frac{0.0190826}{a^2},
\end{equation}
again a remarkably simple and unexpected result.

\section{General right triangular cylinders}\label{sec8}
\label{numerical}
	
As a generalization of the hemiequilateral triangular
cylinder and isosceles right triangular cylinder,
we may consider cylinders having as their cross sections arbitrary right triangles
with hypotenuse $a$ and angles $\theta$, $\pi/2-\theta$, and $\pi/2$. Of course, away from the special
values $\theta=\pi/6$, $\pi/4$, or $\pi/3$, it is not possible to obtain explicit expressions for the
triangle eigenfrequencies $\gamma_n$. Instead, we may employ the efficient scaling method introduced
by Vergini and Saraceno~\cite{vergini} and further developed by Barnett~\cite{barnett}
to obtain numerically all Dirichlet eigenfrequencies $\gamma_n$ for a given triangle with high accuracy, up
to some desired maximum frequency $\gamma_{\rm max}$. The sum over eigenfrequencies is then evaluated numerically
using point-splitting regularization as in Eq.~(\ref{e-ptsplt}),
\be
\mathcal{E}_{\rm num}(\tau)=\frac{1}{2}  
\int_{-\infty}^{\infty} \frac{dk}{2\pi}\,
\sum_{\gamma_n <\gamma_{\rm max}} \sqrt{k^2+ \gamma_{n}^{2}} e^{-\tau \sqrt{k^2+ \gamma_{n}^{2}}},\label{e-num}
\ee
where the error associated with terminating the sum at $\gamma_{\rm max}$ is proportional to $e^{-\tau \gamma_{\rm max}}$,
and thus we must consider $\tau \gamma_{\rm max} \gg 1$.

To extract a physical Casimir energy at small $\tau$, we must first subtract off the known
divergences associated with the first three terms in the Weyl expansion for the spectral density,
associated with the area, perimeter, and corners of the triangular cross section,
\be
\mathcal{E}_{\rm div}(\tau)=\frac{1}{2}  
\int_{-\infty}^{\infty} \frac{dk}{2\pi}\,
\int_0^{\gamma_{\rm max}} d\gamma \sqrt{k^2+ \gamma^{2}} \,e^{-\tau \sqrt{k^2+ \gamma^{2}}}
\left(\frac{A\gamma}{2\pi}-\frac{P}{4\pi}\right)+\frac{C}{48\pi \tau^2} ,
\label{e-div}
\ee
where for the right triangle $A=\frac{a^2}{4} \sin(2\theta)$, $P=a(1+\cos \theta +\sin\theta)$, and the corner
coefficient is given by Eq.~(\ref{cornerterm}), $C=\frac{\pi^2+\pi\theta-2\theta^2}{\theta(\pi-2\theta)}$.
For $\gamma_{\rm max} \to \infty$ at fixed $\tau$, the divergent term recovers the form given earlier
in Eq.~(\ref{div-e}).

We notice that even after subtracting the divergent terms, the limit $\tau \to 0$ may not be taken numerically due to the constraint $\tau\gamma_{\rm max}\gg 1$, imposed by the fact that we have included only a finite part of the eigenfrequency
spectrum. Instead we fix $\gamma_{\rm max}$ and fit
\be
 \mathcal{E}_{\rm num}(\tau) - \mathcal{E}_{\rm div}(\tau)
 = \mathcal{E}+ z_1 \tau+ z_2 \tau^2 \ln \tau +z_3 \tau^2+ \cdots,
\label{fit}
\ee 
over a range $\tau \in[\tau_1,\tau_2]$, where $\gamma_{\rm max}^{-1} \ll \tau_1 < \tau_2 \ll a$,
and where on the right hand side of Eq.~(\ref{fit}) we have listed all non-divergent terms that may
appear for a generic cavity, through order $\tau^2$. The finite term $\mathcal{E}$ is the Casimir energy per unit
length for the cylindrical waveguide of chosen cross section.

\begin{figure}
\includegraphics{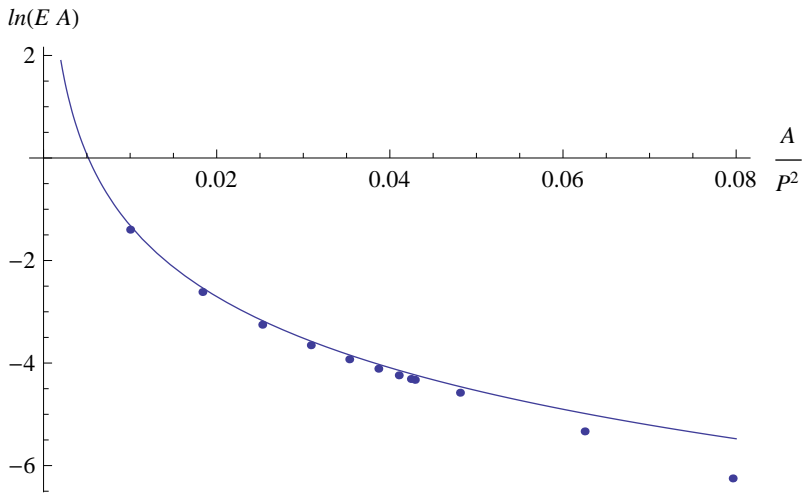}
\caption{\label{fig:pfacyl} 
(Color online)
Graph of the interior Dirichlet Casimir energy per length $\mathcal{E}$
for  various cylinders, multiplied by the area $A$.  
These are plotted as a function of the dimensionless
ratio of the area $A$ to the square of the perimeter  $P$,
$A/P^2$.  The points shown are for the 
$30^\circ$-$60^\circ$-$90^\circ$ triangle,
the isosceles right triangle, the equilateral triangle, the square,
and the circle (which includes interior and exterior contributions),
as well as other right triangles evaluated numerically.
The curve is the PFA approximation, which should become exact
when one of the angles of the triangle goes to zero.}
\end{figure}

For the numerical results shown in Fig.~\ref{fig:pfacyl}, 
we have set $a=1$, $\gamma_{\rm max}=400$, $\tau_1=0.05$, and
$\tau_2=0.2$. With these parameters, the numerical approach reproduces exact results for the hemiequilateral and right
isosceles triangles to better than 1\%. Clearly, the numerical method briefly discussed here is not restricted in
any way to right triangular cylinders or to Dirichlet boundary conditions, but may be applied to arbitrary cavities
in any dimension, provided that a reliable method of numerically evaluating the spectrum is available. A very similar
approach was used in Ref.~\cite{fkw} to evaluate 
the vacuum energy in quantum graphs where no explicit solution is available.
	
The graph shows the comparison of the exact and numerical data to the
proximity force approximation (PFA) for the triangles, in which the energy is computed as
though each opposite element of a right triangle, with the acute angle
$\theta\to0$, has the interaction energy given by that for parallel plates.
This gives the formula
\be
\mathcal{E} A=\frac{\pi^2}{368640}\left(\frac{A}{P^2}\right)^{-2}.
\ee
This result closely matches the numerical data for very acute triangles.

\section{Circular cylinder and divergences}\label{sec9}
\label{circular}
We have considered cylinders of polygonal shapes in
this paper; we believe that we have covered all
cases in which the eigenvalues can be explicitly given.\footnote{We
believe that the analysis in Ref.~\cite{ahmedov} is in error.}
Furthermore, we have given numerical results for Dirichlet
boundary conditions for a variety of right triangles.
It would, of course, be of great interest to consider general
polygons, in particular regular polygons, but that is not
possible analytically with the exception of the equilateral
triangle and the square.  (For example, only some of the eigenmodes
of a hexagon are known.) Instead, one may use the method 
described in Sec.~\ref{sec8}
to obtain good numerical approximations for polygons of arbitrary shape.

Of course, there are well-known analytic results for circular
cylinders \cite{deraad,nesterenko,CaveroPelaez:2006rt,gosdzinsky}.  
In these cases,
however, both interior and exterior contributions to the
cylindrical shell are included, which is necessary,
because otherwise a finite energy cannot be calculated.
Put another way, the $a_2$ heat kernel coefficient proportional
to the cube of the curvature cancels only when interior and
exterior modes are included \cite{Kirsten:2001wz,Fulling:2003zx}.  
If $a_2\ne0$,
the Casimir energy cannot be unambiguously computed, because
an arbitrary logarithmic scale will add to it.

Why does this problem not arise here, where it is impossible
to compute the exterior modes?  Let us consider the heat
kernel for the square, where the mode sum is especially simple.
The heat kernel, which lies at the root of the analytic approach followed 
here,  is expressed in terms of the operator $H=-\nabla^2$
\be
K(t)=\mbox{Tr}\,e^{-Ht}=\int_{-\infty}^\infty \frac{dk}{2\pi}
\sum_{m,n=1}^\infty e^{-(k^2+\gamma_{mn}^2)t}
=\frac12\frac1{\sqrt{\pi t}}\sum_{m,n=1}^\infty e^{-\gamma_{mn}^2t}.
\ee
The Poisson summation formula (\ref{psf}) converts the latter sum into a
form suitable for small $t$ expansion:
\be
\sum_{m=1}^\infty e^{-t m^2}=-\frac12+\frac12\sqrt{\frac\pi t}
+\sqrt{\frac\pi t}\sum_{p=1}^\infty e^{-\pi^2 p^2/t}.
\ee
The latter sum represents exponentially small corrections as
$t\to 0+$, so by squaring this, we get the heat-kernel expansion
\be
K(t)\sim \frac{a^2}{8\pi^{3/2}t^{3/2}}-\frac{a}{4\pi t}+\frac1{8\sqrt{\pi t}},
\ee
where the omitted terms are exponentially small.  The three nonzero heat
kernel coefficients, of course, are proportional to the area, the perimeter,
and a constant (this is the generalized Weyl theorem):
\be
a_0=\frac{A}{8\pi^{3/2}},\quad a_{1/2}=-\frac{P}{16\pi},\quad a_1=\frac1{8\sqrt{\pi}}.
\ee
There is no $a_2$ coefficient.  Carrying out the same calculation for
the equilateral triangular waveguide, we obtain the same result except
that $a_1=1/(6\sqrt{\pi})$.  
These heat kernel coefficients reflect the divergences found in
the energy, Eq.~(\ref{div-e}), as explained in Ref.~\cite{Fulling:2003zx}.  
Somewhat mysteriously, the corner divergences,
which are physically present, appear in the heat-kernel machinery only
in the $a_1$ coefficient, and not in the $a_2$ coefficient. Because
the surfaces are flat, there are no curvature divergences as are present in
the circular case.  In general,
the corner angles $\alpha_i$ appear in the coefficient $a_1$ as
\be
a_1=\frac{C}{48\sqrt{\pi}}=\frac1{48\sqrt{\pi}}\sum_i\left(\frac\pi{\alpha_i}-\frac{\alpha_i}
\pi\right).
\ee
See Eq.~(\ref{cornerterm}) and Refs.~\cite{pleijel,kac,singer}.
Thus, the same remarks apply to the other polygonal shapes considered here.
For a regular $N$-polygonal cross section,
\be
a_1=\frac1{12\sqrt{\pi}}\frac{N-1}{N-2},
\ee
which reduces to the expected result for a circle, $1/(12\sqrt{\pi})$,
as $N\to\infty$.

\section{Plane figures}\label{sec10}
Perhaps it is even more surprising that finite values are obtained for
Casimir energies for polygons in a plane, since the corresponding results
for a circle are divergent, even when interior and exterior contributions
are included \cite{sen,sen1}.  Indeed, Ambj\o rn and Wolfram gave Casimir
energies for rectangles in a plane \cite{ambjorn}. 

The $d=0$ version of the Casimir energy (\ref{CasEDimReg}) is of course
\be
E=\frac12\sum_{m,n}\gamma_{mn},
\ee
which may be immediately evaluated for the four figures considered in
this paper via the Chowla-Selberg formula, or from the Dirichlet $L$-series
formul\ae. The results are given in
Table \ref{tab2}.  (Note there appears to be a transcription error in
the formula given in Ref.~\cite{inui}.)

\begin{table}
\begin{tabular}{ l c c   r}
\hline
Cross section & Dirichlet  & Neumann  &  $A/P^2$ \\
\hline
Hemiequilateral Tr. & $0.174790/h$ & $-0.238159/h$ &  0.038675 \\
Isosceles Right Tr. & $0.113080/a$ & $-0.202939/a$  & 0.042893 \\
Equilateral Tr. & $0.0877806/h$ & $-0.214519/h$  & 0.048113 \\
Square & $0.0410406/a$ & $-0.220759/a$   & 0.0625 \\
\hline
\end{tabular}
\caption{\label{tab2} Casimir energies for plane figures.
The second and third  columns give the Dirichlet and Neumann energies,
expressed in terms of the height $h$ for the figures derived from the
equilateral triangle, and the side $a$ for the figures derived from
the square.
The fourth column gives the dimensionless ratio of the
area to the square of the perimeter of the cross section.  
All results
refer to interior contributions only.}
\end{table}

\begin{figure}
 \includegraphics{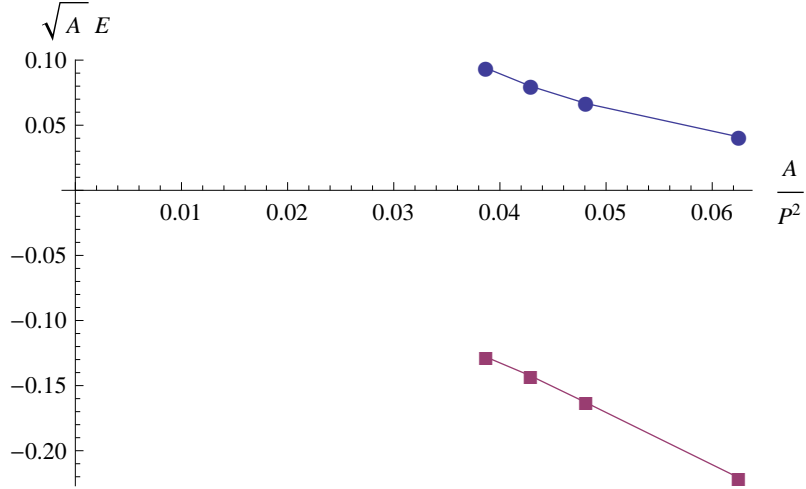}
\caption{\label{fig:pl} 
(Color online)
Graph of the interior Casimir energy $E$
for  various plane figures, multiplied by the
square-root of the area $A$.  
These are plotted as a function of the dimensionless
ratio of the area $A$ to the square of the perimeter  $P$,
$A/P^2$.  The points shown are for the 
$30^\circ$-$60^\circ$-$90^\circ$ triangle,
the isosceles right triangle, the equilateral triangle, and the square.  
The upper curve shows the result for Dirichlet modes, 
the lower curve the energy for Neumann
modes.}
\end{figure}

\begin{figure}
 \includegraphics{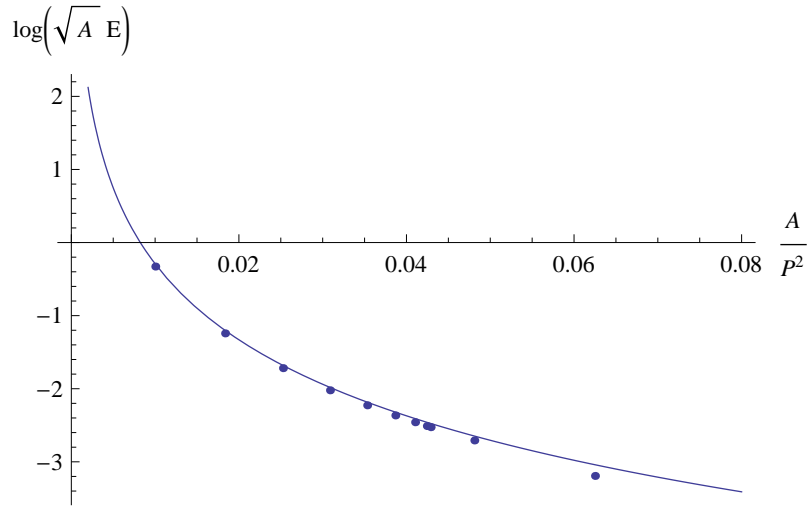}
\caption{\label{fig:pfapl} 
(Color online)
Graph of the interior Dirichlet Casimir energy $E$
for  various plane triangles and the square, multiplied by the 
square root of the area $A$.  
These are plotted as a function of the dimensionless
ratio of the area $A$ to the square of the perimeter  $P$,
$A/P^2$.  The points shown are for the 
$30^\circ$-$60^\circ$-$90^\circ$ triangle,
the isosceles right triangle, the equilateral triangle, and the square,
as well as other right triangles evaluated numerically.
The curve is the PFA approximation, which should become exact
when one of the angles of the triangle goes to zero.}
\end{figure}

These results are plotted versus $A/P^2$ in Fig.~\ref{fig:pl}.
In Fig.~\ref{fig:pfapl} we contrast the PFA, given here by
\be
E\sqrt{A}=\frac{\zeta(3)}{512\pi}\left(\frac{A}{P^2}\right)^{-3/2},
\ee
with the Dirichlet Casimir energies found both analytically, 
and, for general right triangles, numerically
using the method of Sec.~\ref{sec8}.

\section{Analysis and conclusions}\label{sec11}

Our results are summarized in Table \ref{tab1}.
\begin{table}
\begin{tabular}{ l ccc  r}
\hline
Cross section & Dirichlet  & Neumann & EM  & $A/P^2$ \\
\hline
Hemiequilateral Tr. & 0.0163745 & $-0.0204438$ & $-0.00406928$ & 0.038675 \\
Isosceles Right Tr. & 0.0131650 & $-0.0227063$ & $-0.00954135$ & 0.042893 \\
Equilateral Tr. & 0.0102705 & $-0.0265477$ & $-0.0162772$ & 0.048113 \\
Square & 0.00483155 & $-0.0429968$ & $-0.0381653$ & 0.0625 \\
Circle &\textbf{0.00193145} & $\mathbf{-0.0445355}$ & $\mathbf{-0.0426041}$ 
& 0.079577 \\
\hline
\end{tabular}	
\caption{\label{tab1} Casimir energies per unit length 
for cylinders of various cross sections.
The second, third, and fourth columns give the E, H, and EM 
(perfectly conducting) energies/length
multiplied by the cross
sectional area.  The fifth column gives the dimensionless 
ratio of the cross-sectional
area to the square of the perimeter of the cross section.  All results
refer to interior contributions only, with the exception of the final row, 
which gives the energies for a shell of circular cross section 
including both interior and exterior modes.}
\end{table}

\begin{figure}
 \includegraphics{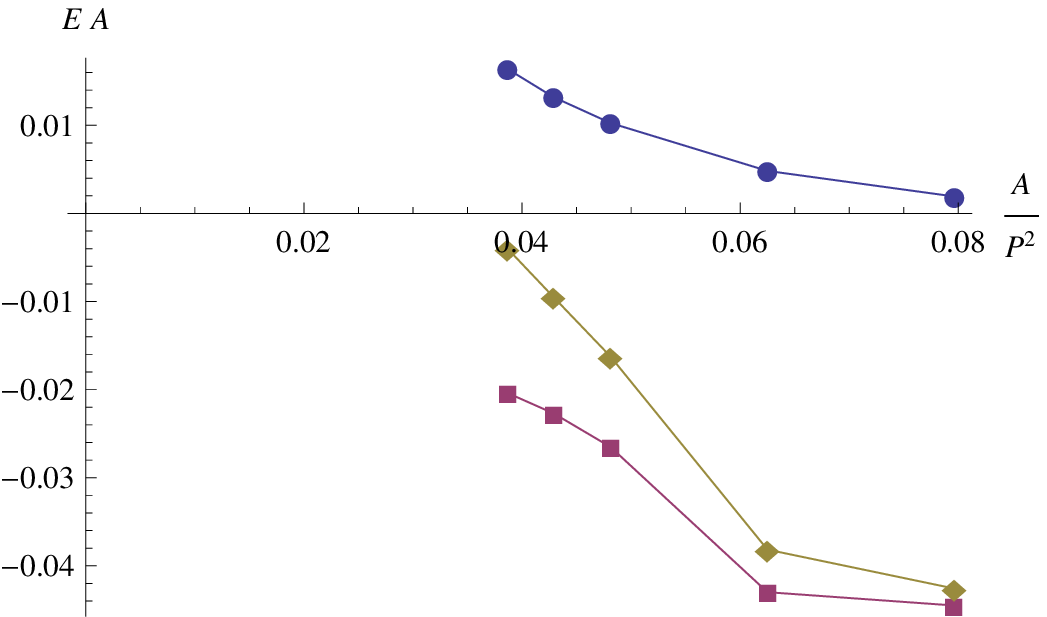}
\caption{\label{fig:uf} (Color online)
Graph of the interior Casimir energy per unit length 
$\mathcal{E}$
for cylindrical waveguides with various cross sections, multiplied by the 
cross-sectional area $A$.  These are plotted as a function of the dimensionless
ratio of the area $A$ to the square of the perimeter of the cross section $P$,
$A/P^2$.  The points shown are for the $30^\circ$-$60^\circ$-$90^\circ$ 
triangle,
the isosceles right triangle, the equilateral triangle, and the square.  
The last
point is for a circle, including  both interior and exterior modes.  The upper
curve shows the energy for E modes (Dirichlet), the lower curve the energy 
for H
modes (Neumann), and the intermediate curve the energy for the sum of the two modes, 
that is, for a perfectly conducting cylinder.}
\end{figure}

We plot the results in Fig.~\ref{fig:uf}. In both the E and H mode, the results
follow a smooth curve for all four polygonal cross sections.  Even the circle
seems to follow the same pattern for the E modes, 
but deviates significantly for
H modes, which is hardly surprising since exterior physics is included for the
circular boundary.  We would speculate that if Casimir energies for
other regular polygonal cross sections could be evaluated numerically, they
would lie on this universal curve.  As we have seen, for the E or Dirichlet
modes, the numerical results for arbitrary right triangles lie on the same
universal curve, which closely approaches that expected from the proximity
force approximation for small angles.

\begin{figure}
 \includegraphics{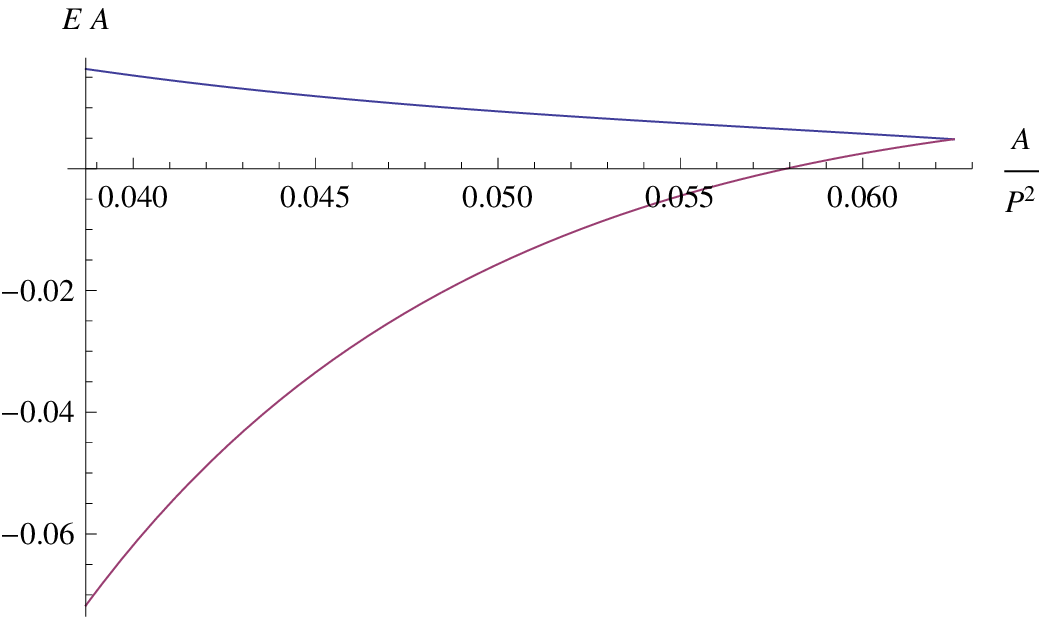}
\caption{\label{fig:rec} (Color online) Graph of the interior Dirichlet 
Casimir energy per unit length $\mathcal{E}$
for cylinders of rectangular cross section, multiplied by the 
cross-sectional area $A$.  This quantity is 
plotted as a function of the dimensionless
ratio of the area $A$ to the square of the perimeter of the cross section $P$,
$A/P^2$. The upper curve is the interpolation of the triangular and
square energies shown in Fig.~\ref{fig:uf}.}
\end{figure}

Of course, not all cross sections will do so, 
as illustrated in Fig.~\ref{fig:rec}.
This shows the trajectory of the Casimir energy for cylinders of rectangular 
cross
section, written as a function of the variable $A/P^2$.  Here, the ratio of the
sides of the rectangles can be written as
\be
\frac{b}a=\frac{1-8\xi}{8\xi}- \frac{\sqrt{\frac1{16}-\xi}}{2\xi},
\ee
where $\xi=A/P^2$, which takes on the maximum value of $\xi=1/16$ for a square.
It will be noted that compared to what was found above for the triangles, the
behavior of the rectangle energies  is very different,
changing from positive to negative at $\xi=0.057902$, 
which corresponds to $b/a=1.74437$.

%
	
	
	


\acknowledgments
We thank the US Department of Energy (grant number
DE-FG02-04ER41305) and the US National Science 
Foundation (grants number PHY-0968492 and PHY-0545390)
 for partial support of this research.  We thank
Steven Armour, Prachi Parashar, Nima Pourtolami, and Jef Wagner
for collaborative assistance, and Steve Fulling for helpful comments.
\appendix
\section{Chowla-Selberg formula\label{appendix}}
Since it is relatively unfamiliar, in this appendix we sketch the derivation
of the Chowla-Selberg formula from the Abel-Plana formula,
\be
\sum_{n=0}^\infty f(n)=\int_0^\infty dn\,f(n)+\frac12f(0)+i\int_0^\infty
dt\frac{f(it)-f(-it)}{e^{2\pi t}-1},
\ee
which requires that $f(z)$ be analytic in the right half-plane.
If $f(z)$ is analytic in the left half-plane, we can derive the
identical formula for $\sum_{n=0}^\infty f(-n).$

Now consider the sum 
\be
S=\sum_{m=-\infty}^\infty\sum_{n=-\infty}^\infty\!\!{}^{\prime\prime} 
\, f(m,n),
\ee
where the double prime means that the single term $m=n=0$ is
omitted. We take 
\be
f(m,n)=(am^2+bmn+cn^2)^{-s},
\ee
where $\mbox{Re}\, s>1$ so the sum exists.
 It is easy to show that
\be
S=2\sum_{m=1}^\infty \sum_{n=0}^\infty f(m,n)+2\sum_{m=1}^\infty
\sum_{n=0}^\infty f(m,-n)-2a^{-s}\zeta(2s)+2c^{-s}\zeta(2s).
\label{sum1}
\ee
We apply the Abel-Plana formula to the two sums in Eq.~(\ref{sum1}),
with the result
\be
S=2 c^{-s}\zeta(2s)+\mbox{II}+\mbox{III},\label{zeta}
\ee
where
\begin{subequations}
\be
\mbox{II}=2\sum_{m=1}^\infty \int_0^\infty dn\left[f(m,n)+f(m,-n)\right],
\ee
and
\be
\mbox{III}=2i\int_0^\infty dt\frac1{e^{2\pi t}-1}\sum_{m=1}^\infty
\left\{f(m,it)-f(m,-it)+f(m,-it)-f(m,it)\right\},
\ee
\end{subequations}
where the different terms have branch points in the 2nd, 3rd, 
1st, and 4th quadrants, respectively.

It is rather easy to evaluate the integral II in terms of the
integral
\be
\int_0^\infty dt (t^2+1)^{-s}=\frac{\sqrt{\pi}\Gamma(s-1/2)}{2\Gamma(s)}.
\ee
The end-point contributions cancel, with the result
\be
\mbox{II}=\sqrt{\pi}\frac{2^{2s}c^{s-1}}{\Delta^{s-1/2}}\zeta(2s-1)
\frac{\Gamma(s-1/2)}{\Gamma(s)},\label{ii}
\ee
where $\Delta=4ac-b^2$.

The final integral III is evaluated by shifting the contour by an
imaginary amount (the shifts along the imaginary axis vanish) and encircling
the branch point, so that we obtain
\bea
\mbox{III}&=&
2i\left(\frac{\sqrt{\Delta}}{2c}\right)^{1-2s}c^{-s}\sum_{m=1}^\infty
m^{1-2s}
\sum_{k=1}^\infty(-2)\int_1^\infty du \left(e^{\pi i s}-e^{-i\pi s}\right)
(u^2-1)^{-s}e^{-\pi \sqrt{\Delta}up/c}\nn\\
&&\qquad\times\left(e^{-imkb\pi/c}+e^{imkb\pi/c}\right)\nn\\
&=& 2^{s+5/2}\pi^s\frac{\Delta^{1/4-s/2}}{\sqrt{c}}\frac1{\Gamma(s)}
\sum_{p=1}^\infty \cos(\pi b p/c)
p^{s-1/2}\sigma_{1-2s}(p)K_{s-1/2}(\pi p\sqrt{\Delta}/c),
\label{iii}
\eea
which uses the representation of the modified Bessel function,
\be
K_\nu(z)=\frac{\sqrt{\pi}(z/2)^\nu}{\Gamma(\nu+1/2)}\int_1^\infty dt\,
(t^2-1)^{\nu-1/2}e^{-zt}.
\ee
Using Eqs.~(\ref{ii}) and (\ref{iii}) in Eq.~(\ref{zeta}), we
find the Chowla-Selberg formula (\ref{cs}) with $a\leftrightarrow c$.



\begin{thebibliography}{99}
\bibitem{casimir}
H. B. G. Casimir, \textit{}Proc. Kon. Ned. Akad. Wetensch. {\bf51}, 793 (1948).


\bibitem{sparnaay} M. Y. Sparnaay, Physica {\bf 24}, 751 (1958).

\bibitem{lamoreaux} S. K. Lamoreaux, Phys.\ Rev.\ Lett.\ {\bf 78}, 5 (1997).

\bibitem{mohideen} U. Mohideen and A. Roy, Phys.\ Rev.\ Lett.\ {\bf81}, 4549
(1998).


\bibitem{Decca:2005yk}
  R.~S.~Decca, D.~Lopez, E.~Fischbach, G.~L.~Klimchitskaya, 
D.~E.~Krause, and V.~M.~Mostepanenko,
  Ann.\ Phys.\ (NY)  {\bf 318}, 37 (2005).

\bibitem{boyer} T. H. Boyer, Phys.\ Rev.\ {\bf 174}, 1764 (1968).

\bibitem{casimir56} H. B. G. Casimir, Physica {\bf 19}, 846 (1956).

\bibitem{lukosz}W. Lukosz, \textit{}Physica {\bf56}, 109 (1971).

\bibitem{ambjorn}
J. Ambj{\o}rn and S. Wolfram, \textit{}Ann. Phys. (NY) {\bf147}, 1 (1983).

\bibitem{deraad}
L. L. DeRaad, Jr. and K. A. Milton, 
\textit{}Ann. Phys. (NY) {\bf136}, 229 (1981).

\bibitem{gosdzinsky}
P. Gosdzinsky and A. Romeo, \textit{} Phys. Lett. B {\bf441}, 265 (1998).




\bibitem{elizalde}
E. Elizalde, S. D. Odintsov, A. Romeo, A. A. Bytsenko, and S. Zerbini, 
\textit{Zeta Regularization Techniques with Applications} (World Scientific, 
1994).

\bibitem{cs}
S. Chowla and A. Selberg, 
J. reine u. angewandte Math. {\bf227}, 86 (1967).
 
\bibitem{lerch}
M. Lerch, 
Bull.\ sci.\ math.\ {\bf21}, 290 (1897).

\bibitem{lorenz}
L. Lorenz, Matematisk Tidsskrift {\bf1}, 97 (1871).

\bibitem{hardy}
G. H. Hardy, Messenger Math.\ {\bf49}, 85 (1919).

\bibitem{fletcher} A. Fletcher, J. C. P. Miller, L. Rosenhead, and L. J.
Comrie, {\it An Index of Mathematical Tables\/} (Blackwell, London, 1962),
Vol.~1, p.~95.

\bibitem{zucker} I. J. Zucker, J. Math.\ Phys.\ {\bf 15}, 187 (1974).

\bibitem{gz}
M. L. Glasser and I. J. Zucker, {\it Theoretical Chemistry: Advances
and Perspectives\/} (Academic, New York, 1980), Vol.~5, p.~67.

\bibitem{itzykson} C. Itzykson and J. M. Luck, J. Phys. A {\bf 19}, 211
(1986).

\bibitem{itzykson2} C. Itzykson, P. Moussa, and J. M. Luck, J. Phys.\ A
{\bf19}, L111 (1986).

\bibitem{kvitsinsky} A. A. Kvitsinsky, J. Phys.\ A {\bf 29}, 6379 (1996).

\bibitem{lukosz2} W. Lukosz, Z. Phys.\ {\bf 258}, 99 (1973).

\bibitem{lukosz3} W. Lukosz, Z. Phys.\ {\bf 262}, 327 (1973).

\bibitem{zimerman} J. R. Ruggiero, A. H. Zimerman, and A. Villani,
Rev.\ Bras.\ Fis.\ {\bf 7}, 663 (1977).
\bibitem{zimerman2} J. R. Ruggiero, A. H. Zimerman, and A. Villani,
J. Phys.\ A  {\bf 13}, 761 (1980).

\bibitem{nesterenko} K. A. Milton, A. V. Nesterenko, and V. V.
Nesterenko, Phys.\ Rev.\ D{\bf 59} 105009 (1999).

\bibitem{inui}N. Inui, \textit{}J. Phys. Soc. Jap. {\bf 76}, 11 (2007).




\bibitem{ahmedov}H. Ahmedov and I. H. Duru, \textit{}J. Math. Phys. {\bf45}, 
3 (2004).

\bibitem{hazlett}  R. D. Hazlett and D. K. Babu, Quart.\ Appl.\ Math.\
{\bf 67}, 579 (2009).

\bibitem{alvarez} E. \'Alvarez, F. D. Mazzitelli, A. G. Monastra, and D. A.
Wisniacki, arXiv:1007.4742.



\bibitem{radbook}
K. A. Milton and J. Schwinger, \textit{Electromagnetic Radiation: 
Variational Methods, Waveguides and Accelerators} (Springer, 2006).

\bibitem{embook}
J. Schwinger, L. L. DeRaad, Jr., K. A. Milton, and W.-y. Tsai, 
\textit{Classical Electrodynamics} (Westview Press, 1998).


\bibitem{Milton:2010qr}
  K.~A.~Milton,
  arXiv:1005.0031 [hep-th].


\bibitem{stratton} J. A. Stratton, \textit{Electromagnetic Theory}
(McGraw-Hill, New York, 1941).

\bibitem{lame} M. G. Lam\'e, {\it Le\c{c}ons sur la th\'{e}orie 
math\'{e}matique
de l'\'{e}lasticit\'e\ des corps solides} (Bachelier, Paris, 1852).

\bibitem{vergini} E. Vergini and M. Saraceno, Phys. Rev. E {\bf 52},
2204 (1995).

\bibitem{barnett} A. H. Barnett, Comm. Pure Appl. Math. {\bf 59}, 1457 (2006). 	 

\bibitem{fkw} S. A. Fulling, L. Kaplan, and J. H. Wilson, Phys. Rev. A
{\bf 76}, 012118 (2007). 

\bibitem{CaveroPelaez:2006rt}
  I.~Cavero-Pel\'aez, K.~A.~Milton, and K.~Kirsten,
  J.\ Phys.\ A  {\bf 40}, 3607 (2007).

\bibitem{Kirsten:2001wz}
  K.~Kirsten,
{\it Spectral Functions in Mathematics and Physics}
(Chapman \& Hall/CRC, Boca Raton, FL, 2001).


\bibitem{Fulling:2003zx}
  S.~A.~Fulling,
  J.\ Phys.\ A  {\bf 36}, 6857 (2003).

\bibitem{pleijel} A. Pleijel, Ark.\ Mat.\ {\bf 2}, 663 (1954).

\bibitem{kac} M. Kac, Am.\ Math.\ Mon.\  {\bf 73}, 1 (1966).

\bibitem{singer} H. P. McKean and I. M. Singer, J. Diff.\ Geom.\
{\bf 1}, 43 (1967).



\bibitem{sen} S. Sen, Phys.\ Rev.\ D {\bf 24}, 869 (1981).

\bibitem{sen1} S. Sen, J. Math.\ Phys.\ {\bf 22}, 2968 (1981).



\end{thebibliography}
\end{document}